# THE DYNAMICS OF ASTEROID ROTATION, GOVERNED BY YORP EFFECT: THE KINEMATIC ANSATZ.

**Sergey V. Ershkov**

Sternberg Astronomical Institute,

M.V. Lomonosov's Moscow State University,

13 Universitetskij prospect, Moscow 119992, Russia

e-mail: sergej-ershkov@yandex.ru

**Roman V. Shamin**

Moscow Technological University (MIREA),

78 Vernadsky Avenue, Moscow 119454,

e-mail: roman@shamin.ru

The main motivation of this research is the analytical exploration of the dynamics of asteroid rotation when it moves in elliptic orbit through Space.

According to the results of *Efroimsky, Frouard* (2016), various perturbations (collisions, close encounters, YORP effect) destabilize the rotation of a small body (asteroid), deviating it from the initial-current spin state. This yields evolution of the spin towards rotation about maximal-inertia axis due to the process of *nutation relaxation* or to the proper spin state corresponding to minimal energy with a fixed angular momentum.

We consider in our research the aforementioned spin state of asteroid but additionally under non-vanishing influence of the effects of non-gravitational nature (YORP effect), which is destabilizing the asteroid rotation during its motion far from giant planets.

Meanwhile, new solutions for asteroid rotation dynamics in case of negligible (time-dependent) applied torques have been obtained in our development. New method for solving Euler's equations for rigid body rotation is suggested; an elegant example for evolution of spin towards the rotation about maximal-inertia axis is calculated.

# 1. Introduction, the system of equations.

Motion of asteroids is known to be under the influence of effects of non-gravitational nature (Yarkovsky effect, YORP effect). The Yarkovsky effect is the force acting on a rotating body in space caused by the anisotropic emission of thermal photons, which carry momentum [1]. It is usually considered in relation to meteoroids or small asteroids (about 10 cm to 10 km in diameter), as its influence is most significant for these bodies. Such a force is produced by the way an asteroid absorbs energy from the sun and re-radiates it into space as heat by anisotropy.

In fact, there exists a disbalance of momentum when asteroid at first absorbs the light radiating from the sun, but then asteroid re-radiates the heat. Such a disbalance is caused by the rotation of an asteroid during period of warming as well as by the anisotropic cooling of surface and inner layers; the processes above depend on anisotropic heat transfer in the inner layers of asteroid.

During decades such a disbalance forms a negligible, but important additional acceleration for small bodies due to Yarkovsky effect. E.g. the asteroid RQ36 wandered off course 160 km in 12 years due to heat radiating from its surface. Thus, Yarkovsky effect is small but very important in celestial mechanics as well as in calculating of the proper orbits of asteroids and other small bodies.

Besides, Yarkovsky effect is not predictable (it could be only observed & measured by astronomical methods); the main reason is the unpredictable character of the rotating of small bodies [2], even in the case when there is no any collision between them.

If regime of rotation of asteroid is changing, we could observe a generalization of Yarkovsky effect, i.e. the Yarkovsky–O'Keefe–Radzievskii–Paddack effect or YORP effect [2]. We should especially note that due to YORP effect, a physical disintegration of asteroids is possible due to self-destruction under the influence of sudden acceleration during a fast rotation. For example, asteroid "P/2013 R3" was observed breaking apart in 2013 perhaps due to the YORP effect [3].

Difference between the aforementioned Yarkovsky and the YORP effects comes from the common knowledge [1-2] and seems very trivial: they both originate from recoil light pressure forces, but the Yarkovsky drag is a force (which is applied to the center of mass of the asteroid, and thus altering its orbital dynamics), while the YORP is a torque, thus altering the rotation of the asteroid around its center of mass. We should especially note that *Yarkovsky* effect and YORP effect [1-2], which are assumed to be the effects of non-gravitational nature, are manifested to be considerable for sizes of small bodies less than < 10 km.

The main motivation of this research is the analytical exploration of the dynamics of asteroid rotation when it moves in elliptic orbit through Space.

Based on the assumption of asteroid rotating as rigid body (it means distances between various points inside the rigid body should be preferably constant or should be elongated negligibly), let us recall that Euler equations describe rotation of rigid body in a frame of reference fixed in the rotating body [4-5]:

$$I_1 \frac{d\Omega_1}{dt} + (I_3 - I_2)\cdot\Omega_2\cdot\Omega_3 = K_1,$$

$$I_2 \frac{d\Omega_2}{dt} + (I_1 - I_3)\cdot\Omega_1\cdot\Omega_3 = K_2, \qquad (1)$$

$$I_3 \frac{d\Omega_3}{dt} + (I_2 - I_1)\cdot\Omega_1\cdot\Omega_2 = K_3,$$

where $K_i = K_i(t)$ are applied torques, $I_i \neq 0$ are principal moments of inertia, and $\Omega_i$ are components of angular velocity vector along the principal axes (i = 1, 2, 3).

Let us present aforementioned system (1) in another form: if we multiply each appropriate equation of system (1) on $\Omega_i$ properly

$$I_1\Omega_1 \frac{d\Omega_1}{dt} + (I_3 - I_2)\cdot\Omega_1\cdot\Omega_2\cdot\Omega_3 = K_1\cdot\Omega_1,$$

$$I_2\Omega_2 \frac{d\Omega_2}{dt} + (I_1 - I_3)\cdot\Omega_1\cdot\Omega_2\cdot\Omega_3 = K_2\cdot\Omega_2, \qquad (2)$$

$$I_3\Omega_3 \frac{d\Omega_3}{dt} + (I_2 - I_1)\cdot\Omega_1\cdot\Omega_2\cdot\Omega_3 = K_3\cdot\Omega_3,$$

we should obtain in result

$$I_i \Omega_i \frac{d\Omega_i}{dt} = K_i\cdot\Omega_i - C_i\cdot f(t), \qquad (3)$$

$$\left\{ C_1 = (I_3 - I_2),\ \ C_2 = (I_1 - I_3),\ \ C_3 = (I_2 - I_1),\ \ f(t) = \Omega_1\cdot\Omega_2\cdot\Omega_3 \right\}$$

Eqn. (3) does not give an *Abel* ODE of the 2-nd type [6] (a kind of generalization of *Riccati* ODE), but rather a system of 3 *Abel*-type ODEs, which is substantially more complicated. There is no obvious gain in using Eqn. (3) instead of Eqn. (1): they are both (presumably unsolvable) systems of three non-linear ODEs. As for the *Riccati*-type ODE, we should additionally note that a modern methods exist for obtaining the solution of *Riccati* equations with a good approximation [7-8].

But in order to confirm the actual *Abel* character of the solutions of Eqs. (1), let us express the component $\Omega_3$ of a solution from the second equation of system (1), for example (for definiteness, $I_1 \geq I_2 \geq I_3, I_1 \neq I_3$):

$$\Omega_3 = \frac{K_2 - I_2 \frac{d\Omega_2}{dt}}{(I_1 - I_3)\cdot \Omega_1}, \qquad (4)$$

which should be substituted as $\Omega_3$ in the first equation of (1)

$$I_1(I_1 - I_3)\cdot \Omega_1 \cdot \frac{d\Omega_1}{dt} = K_1(I_1 - I_3)\cdot \Omega_1 + \left\{(I_2 - I_3)\cdot \Omega_2 \cdot \left(K_2 - I_2 \frac{d\Omega_2}{dt}\right)\right\}, \qquad (5)$$

As we can see, Eqn. (5) is also *Abel* ODE of 2-nd type in regard to the function $\Omega_1$; the aforementioned equation reveals that the component $\Omega_1$ non-linearly depends on component $\Omega_2$, and *vice versa*.

Meanwhile, we should additionally note that for the reason of a special character of the solutions of *Riccati*-type ODE [6], there exists a possibility for sudden *jumping* of the magnitude of the solution at some meaning of time-parameter *t*.

In the physical sense, such jumping of the *Riccati*-type solutions could be associated with the effect of sudden acceleration of rigid body rotation around the appropriate principle axis at definite moment of parametric time $t_0$.

2. **<u>Refering to the results of Precession Relaxation of Viscoelastic Oblate Rotators.</u>**

According to the results [9], various perturbations (collisions, close encounters, YORP effect [10]) destabilize the rotation of asteroid, deviating it from the initial-current spin state. But the body is known to be experiencing the additional stress, generated by inertial forces, for example, during the motion along appropriate orbit through Space near giant planets or other massive objects. So, ensuing inelastic dissipation [11-17] (tidal dissipation and energy dissipation on internal friction for bodies in non-principal rotation state) reduces kinetic energy, without influencing angular momentum. This yields evolution of spin towards rotation about maximal-inertia axis [9] due to process of *nutation relaxation* or to the proper spin state corresponding to minimal energy with a fixed angular momentum.

Let us explore the final dynamical state of asteroid rotation, ignoring process of *nutation relaxation* which was successfully fully explored in [9]. Asteroid is supposed to be moving along its orbit far from the close influences of additional gravitational forces or far from *Hill sphere*.

Let us also assume that all the external torques, associated with inertial forces, tides, are neglected in (1), except the effects of non-gravitational nature, which are nevertheless supposed to be negligible enough. Note that if we sum all the equations in (2) (let us sum the left parts of all equations to each other against summing of all right parts of equations (2)), we should obtain

$$\frac{d}{dt}\left(\frac{I_1\Omega_1^2 + I_2\Omega_2^2 + I_3\Omega_3^2}{2}\right) = Y_1 \cdot \Omega_1 + Y_2 \cdot \Omega_2 + Y_3 \cdot \Omega_3, \qquad (6)$$

where we designate $\{Y_1, Y_2, Y_3\}$ to be components of applied torques $K_i$ in (2) due to effect of non-gravitational nature (YORP effect), $\{Y_1, Y_2, Y_3\} \to \{0, 0, 0\}$.

The right part of Eq. (6) could be represented also as below

$$(\vec{Y} \cdot \vec{\Omega}) = Y_1 \cdot \Omega_1 + Y_2 \cdot \Omega_2 + Y_3 \cdot \Omega_3, \qquad (7)$$

so, we obtain from (7) that if angular velocity $\mathbf{\Omega} = \{\Omega_1, \Omega_2, \Omega_3\}$ is supposed to be almost perpendicular to vector of YORP effect, $\mathbf{Y} = \{Y_1, Y_2, Y_3\}$, the inelastic (tidal) dissipation of kinetic energy should be minimal or could be considered as negligible.

But according to results of [9], inelastic (mainly tidal) dissipation, which is reducing kinetic energy, yields evolution of spin towards rotation about maximal-inertia axis with rate of rotation $\Omega_1$ (we have chosen $I_1 \geq I_2 \geq I_3$); it means:

$$\{\Omega_2, \Omega_3\} \ll \Omega_1 \qquad (*)$$

So, the rotation of asteroid about the maximal-inertia axis $I_1$ with rate of rotation $\Omega_1$ should be preferably perpendicular to the vector of total YORP effect for the minimal regime of the inelastic tidal dissipation, which is reducing the kinetic energy of asteroid. In this case, we could obtain (as first approximation) from (3):

$$\{\Omega_2, \Omega_3\} << \Omega_1 \Rightarrow f(t) = \Omega_1 \cdot \Omega_2 \cdot \Omega_3 \to 0, \Rightarrow Y_1 \cong I_1 \frac{d\Omega_1}{dt} \qquad (8)$$

YORP is a torque, acting upon the asteroid and thus causing the angular acceleration. But stating the opposite (that it is the angular acceleration that causes the torque) could also make sense from some profoundly philosophical point of view: indeed, YORP is a torque, causing the angular acceleration of the asteroid (8), and *vice versa*.

### 3. Analytical solution of the initial system of equations.

Let us assume that *non-inertial* forces of non-gravitational nature will manifest the resulting torque $\{Y_1, Y_2, Y_3\}$ on each principal axis ($I_1 \geq I_2 \geq I_3$, $I_1 \neq I_3$).

All in all, we could substitute the expression (4) for $\Omega_3$ in the third equation of system (1) ($\Omega_3 \neq$ const):

$$\frac{I_3}{(I_1 - I_3)} \cdot \left( \frac{(Y_2' - I_2 \frac{d^2 \Omega_2}{dt^2}) \cdot \Omega_1 - (Y_2 - I_2 \frac{d\Omega_2}{dt}) \cdot \frac{d\Omega_1}{dt}}{\Omega_1^2} \right) + (I_2 - I_1) \cdot \Omega_1 \cdot \Omega_2 = Y_3, \quad (9) \Rightarrow$$

then, by expressing $\{\Omega_1 \cdot (d\Omega_1/dt)\}$ from (5), we obtain from equation (9):

$$A \cdot \Omega_1^4 - B \cdot \Omega_1^3 + C \cdot \Omega_1^2 - D \cdot \Omega_1 - E = 0, \qquad (10)$$

$$A = \left( \frac{I_1(I_1 - I_3)^2 \cdot (I_2 - I_1)}{I_3} \cdot \Omega_2 \right), \quad B = \left( \frac{I_1(I_1 - I_3)^2}{I_3} \cdot Y_3 \right), \quad C = I_1(I_1 - I_3) \cdot (Y_2' - I_2 \frac{d^2 \Omega_2}{dt^2}),$$

$$D = (Y_2 - I_2 \frac{d\Omega_2}{dt}) \cdot Y_1 \cdot (I_1 - I_3), \quad E = \left\{ (I_2 - I_3) \cdot \Omega_2 \cdot (Y_2 - I_2 \frac{d\Omega_2}{dt})^2 \right\}$$

Having solved the polynomial equation (10) (of degree 4) in regard to the component of angular velocity $\Omega_1 = \Omega_1 (\Omega_2, \Omega_2', \Omega_2'')$, we should substitute one of the obtained *real* roots the appropriate expression for $\Omega_1$ in equation (5).

Obviously, we could assume at solving such the resulting monstrous ODE of the 3-d order in regard to component of angular velocity $\Omega_2(t)$, consisting of the roots ensuing from Eqn. (10) and their derivatives, that it could not be solved analytically but only by numerical methods.

But thanks to the negligible magnitude of the YORP effect, we could assume at solving of *algebraic* equation (10) that all the magnitudes of the components of YORP tend to zero $\{Y_1, Y_2, Y_3\} \rightarrow \{0, 0, 0\}$; so, we obtain from (10)

$$A \cdot \Omega_1^4 + C \cdot \Omega_1^2 - E = 0, \qquad (11)$$

$$A = \left( \frac{I_1(I_1 - I_3)^2 \cdot (I_2 - I_1)}{I_3} \cdot \Omega_2 \right), \quad B = 0, \quad C = I_1(I_1 - I_3) \cdot (Y_2' - I_2 \frac{d^2 \Omega_2}{dt^2}),$$

$$D = 0, \quad E = \left\{ (I_2 - I_3) \cdot \Omega_2 \cdot (I_2 \frac{d\Omega_2}{dt})^2 \right\}$$

We can see from the structure of biquadratic Eqn. (11) that we still need additional simplifying assumptions to solve it easily. So, let us consider the case of symmetric rigid body rotation, $I_2 \cong I_3$. We should note that YORP effect vanishes for simple shape models [18]-[19] (such as ellipsoids of rotation).

So, we could consider only the case $I_2 \cong I_3$ (but, nevertheless, $I_2 \neq I_3$) for approximate solutions. Meanwhile, Eqn. (11) could be easily solved in this case:

$$A \cdot \Omega_1^4 + C \cdot \Omega_1^2 = 0, \quad \Rightarrow \quad \Omega_1 = \pm \sqrt{-\frac{C}{A}} \quad \left( (Y_2' - I_2 \frac{d^2 \Omega_2}{dt^2}) \geq 0 \right) \qquad (12)$$

$$A = \left( \frac{I_1(I_1 - I_3)^2 \cdot (I_2 - I_1)}{I_3} \cdot \Omega_2 \right), \quad B = 0, \quad C = I_1(I_1 - I_3) \cdot (Y_2' - I_2 \frac{d^2 \Omega_2}{dt^2}),$$

$$D = 0, \quad E = 0$$

We do not consider here the symple case of assuming for all the components of effect of non-gravitational nature $\{Y_1, Y_2, Y_3\} \rightarrow \{0, 0, 0\}$ during solving the initial system (1), for the reason that such symplification should reduce system (1) to the well-known case $\Omega_3 = C = $ const (*the case of harmonic oscillations*), see [9]:

$$I_1 \frac{d\Omega_1}{dt} + (I_3 - I_2) \cdot \Omega_2 \cdot C = 0,$$

$$I_2 \frac{d\Omega_2}{dt} + (I_1 - I_3) \cdot \Omega_1 \cdot C = 0, \qquad (13)$$

$$\Omega_3 = C.$$

So, excluding the case of *harmonic oscillations* (13) (it means that $\Omega_3 \neq$ const in our derivation), we could substitute appropriate expression for $\Omega_1$ from Eqn. (5) the expression for $\Omega_1$ in (12), and *vice versa* ($I_2 \cong I_3$):

$$\Omega_1 = \Omega_1(0) + \frac{1}{I_1}\int (Y_1)\,dt, \;\Rightarrow\; \Omega_1^2 = \frac{I_3 \cdot (Y_2' - I_3 \frac{d^2\Omega_2}{dt^2})}{(I_1 - I_3)^2 \cdot \Omega_2} \;\Rightarrow$$

$$I_3 \frac{d^2\Omega_2}{dt^2} + \left(\frac{(I_1 - I_3)^2}{I_3}\Omega_1^2\right) \cdot \Omega_2 - Y_2' = 0 \qquad (14)$$

$$\left((Y_2' - I_3 \frac{d^2\Omega_2}{dt^2}) \geq 0\right)$$

## **Discussion**

The main motivation of this research is the analytical exploration of dynamics of asteroid rotation when it moves in elliptical orbit through Space, assuming additional influence of *negligible* arbitrary function *Y*(*t*) as effect of non-gravitational nature (YORP effect). Asteroid is supposed to be moving along its orbit far from *Hill sphere* [20] ($a_p$ is semimajor axis of the planet):

$$r_H = a_p \cdot \left(\frac{m_{planet}}{3M_{Sun}}\right)^{\frac{1}{3}}$$

Thus, a kind of *kinematic* ansatz for the investigating of the influence of effect *Y*(*t*) → *0* of non-gravitational nature on asteroid rotation is presented here. The *kinematic* ansatz means that instead of considering the dynamical causes [18-19] for the YORP effect, we should investigate its kinematic influence on elements of orbit as well as its influence on the spin evolution during the process of asteroid rotation.

Let us discuss the application to the real cases showing the actual magnitudes of the considered effects (influences of gravitational torques, YORP effect, etc.).

In a comprehensive article [18], the authors in their results reported precise optical photometric observations of a small near-Earth asteroid, (54509) 2000 PH5, acquired earlier. The small near-Earth asteroid (54509) 2000 PH5 is one of the few known Earth co-orbitals in a near 1:1 mean-motion resonance with Earth. The small radar-derived mean radius of circa 57 m as well as the fast spin period of 12.17 min makes it a very practical target for observations from Earth-based telescopes. During the period of observations, the asteroid has been continuously increasing its rotation rate ω by dω/dt = 2.0 (±0.2) × $10^{-4}$ degrees per $day^{-2}$. The authors also simulated asteroid's close Earth approaches from 2001 to 2005, showing that gravitational torques cannot explain the observed spin rate increase. The obtained result suggests that it would take circa 550,000 years for YORP to double the rotation rate of asteroid 2000 PH5 in the future. From this value we may expect that YORP will cause structural changes, mass shedding, or even fission of this object at some point in the future, depending on its internal strength. This result also implies the possible existence of a population of 100-m asteroids with rotation periods of circa 20 s, significantly faster than the most rapidly rotating asteroid of this size, 2000 WH10 with P of circa 80 s.

The actual comparison with the other well-known asteroids in regard to the various physical properties has been made in [19] (rotation rate change, absolute magnitude, rotation period, obliquity and the solar flux weighted mean heliocentric distance, with semimajor axis *a* and eccentricity *e*). Most of the asteroids appear to have a slow rotation rate (rotation periods ranging from a fraction of an hour to more than 12 hours).

We should especially note that due to YORP effect, a physical disintegration of asteroids is possible due to self-destruction under the influence of sudden accelerating during a fast rotation. For example, asteroid "P/2013 R3" was observed breaking apart in 2013 perhaps due to the YORP effect [3].

The last but not least, we have not discussed or considered here a long-term change ensuing from YORP effect for the obliquity γ of the axis of rotation with respect to orbit of asteroid (see the formulae (5)-(6) in [19]).

## **Conclusion**

YORP effect among various other perturbations destabilizes the rotation of a small body (asteroid), deviating it from the initial-current spin state. But the body is known to be experiencing the additional stress, generated by the inertial forces during the motion along the appropriate orbit through Space near the massive planets. So, the ensuing inelastic (tidal) dissipation reduces the kinetic energy, without influencing the angular momentum. This yields evolution of spin towards the rotation about maximal-inertia axis due to the process of *nutation relaxation* or to the proper spin state corresponding to minimal energy with a fixed angular momentum. But even at this final spin state, the changing of regimes of rotations about the minimum-inertia axes may disrupt the body of asteroid: indeed, as we can see from the development above, a physical disintegration of the asteroids is possible due to self-destruction under the influence of sudden acceleration during a fast rotation.

We have explored here the dynamics of asteroid rotation, considering the final spin state of rotation for a small body (asteroid). We succeeded in obtaining the analytical algorithm for solving the Euler's equations of rigid body rotations with non-zero (but negligible) torques. The proper approximate solutions have been obtained as a result. Solving the resulting polynomial invariant of the ODE-system of rigid body rotation, we assume for the components of YORP $\{Y_1, Y_2, Y_3\} \rightarrow \{0, 0, 0\}$. Such a simplification allows us to reduce the polynomial equation of degree 4 to a bi-quadratic equation; then, additionally assuming the case of symmetric rigid body rotation $I_2 \cong I_3$ (with YORP effect vanishing for ellipsoids of rotation [19]), we have obtained in (14) the components of angular velocity rotation for asteroid as pointed below:

$$\Omega_1 = \Omega_1(0) + \frac{1}{I_1}\int (Y_1)\,dt,$$

$$I_3\frac{d^2\Omega_2}{dt^2} + \left(\frac{(I_1-I_3)^2}{I_3}\Omega_1^2\right)\cdot\Omega_2 - Y_2' = 0, \quad (15)$$

$$\Omega_3 = \frac{Y_2 - I_3\frac{d\Omega_2}{dt}}{(I_1-I_3)\cdot\Omega_1}.$$

We should note that if we consider the case of YORP effect such as $\{Y_1 = \text{const}, Y_2 = \text{const}, Y_3 = \text{const}\} \to \{0, 0, 0\}$, solution (15) could be reduced accordingly

$$\Omega_1 = \Omega_1(0) + \frac{Y_1}{I_1}(t - t_0),$$

$$I_3\frac{d^2\Omega_2}{dt^2} + \left(\frac{(I_1-I_3)^2}{I_3}\Omega_1^2\right)\cdot\Omega_2 = 0 \quad (16)$$

$$\Omega_3 = -\frac{I_3\frac{d\Omega_2}{dt}}{(I_1-I_3)\cdot\Omega_1}$$

where the 2-nd equation of Eqs. (16) for the dynamics of component $\Omega_2$ of angular velocity

$$I_3\frac{d^2\Omega_2}{dt^2} + \left(\frac{(I_1-I_3)^2}{I_3}\Omega_1^2\right)\cdot\Omega_2 = 0 \quad (17)$$

is known to be the *Riccati* ODE [6]. If we consider the short and middle time-scale (with $Y_1 \cong 0$) $\to \Omega_1 \cong \Omega_1(0)$, Eqn. (17) could be considered to have been reduced to the case of *harmonic oscillations*, see [9].

At the long time-scale, we should additionally note that for reason of a special character of the solutions of *Riccati*-type ODE, there exists a possibility for sudden *jumping* of magnitude of the solution at some meaning of time-parameter *t* [21-27].

In the physical sense, such jumping of *Riccati*-type solutions of Eqn. (17) could be associated with effect of sudden acceleration/deceleration of rigid body rotation around appropriate principle axis at definite moment of parametric time $t_0$.

Mathematical procedure of presenting the components of angular velocity (15)-(17) via Euler angles (and Wisdom angles) has been moved to the Appendix, with only basic formulae left in the main text. Elegant examples of approximate solutions have been presented, which obviously are demonstrating the sudden character of *jumping* of the solution's magnitude at periodic ranges of time-parameter *t*.

## **Conflict of interest**

Authors declare that there is no conflict of interests regarding publication of article.

## **Acknowledgements**

## **Appendix *A*1. The dynamics of the asteroid rotation via *Euler* angles.**

Let us define the *fixed* cartesian system of coordinates for convinient representation of asteroid rotation [20] during its motion in the elliptic orbit through Space:

- *Oz* axis is chosen to be perpendicular to the orbit plane of asteroid;
- *Ox* axis is chosen to be coinciding to the velocity vector of asteroid at pericenter of its elliptic orbit;
- *Oy* axis is chosen to be accomplishing the right-handed cartesian coordinate system.

Recall that principle axes, corresponding to principle moments of inertia in a frame of reference fixed in the rotating body (1), are also chosen to be accomplishing the right-handed cartesian coordinate system *Ox′*, *Oy′*, *Oz′*, respectively.

We should transform the components of angular velocity rotation of asteroid (which had been previously resolved (15)-(16) in a frame of reference fixed in the rotating body $Ox'$, $Oy'$, $Oz'$) to the rotations in the *fixed* cartesian system of coordinates $Ox$, $Oy$, $Oz$, determined by the *kinematic* equations of Euler angles [28]:

$$\Omega_1 = \sin\psi \cdot \sin\theta \cdot \frac{d\varphi}{dt} + \cos\psi \cdot \frac{d\theta}{dt},$$

$$\Omega_2 = \cos\psi \cdot \sin\theta \cdot \frac{d\varphi}{dt} - \sin\psi \cdot \frac{d\theta}{dt}, \qquad (18)$$

$$\Omega_3 = \cos\theta \cdot \frac{d\varphi}{dt} + \frac{d\psi}{dt},$$

here notation of Euler angles is the same as that of [28], see Fig.1: $\varphi$ is the angle corresponding to the first rotation about the axis $Oz$ by an angle $\varphi$, $\theta$ is the angle corresponding to the second rotation about the former $Ox$-axis (now $Ow$) by an angle $\theta \in (0, \pi)$, and $\psi$ is the angle corresponding to the last third rotation about axis $Oz'$ (by an angle $\psi$). Notation of Euler angles differs from that of [29]: $\theta$ angle is the same in both notations, but angle $\varphi$ is changed to the angle $\psi$ (and *vice versa*).

Similar system of Euler angles had been employed earlier in [29]-[30], but it differs from the current notation: there $\psi$ angle is the same in both notations, but angle $\varphi$ is changed to the angle $\theta$ (and *vice versa*).

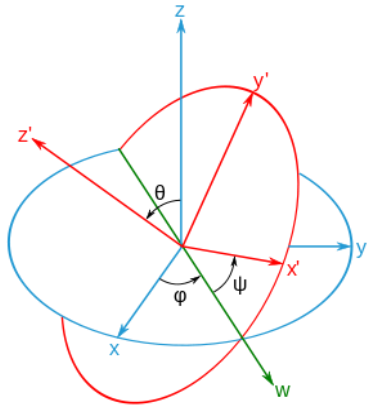

Fig. 1. Notation of Euler angles.

The 1-st and 2-nd equations of system (18) let us obtain (here below $\theta \in (0, \pi)$, $\varphi \neq$ const, $\psi \neq \pi \cdot (2n+1)/2$, $n = 0, 1, 2, 3, \ldots$):

$$\cot \psi = \frac{\Omega_2 + \sin \psi \cdot \frac{d\theta}{dt}}{\Omega_1 - \cos \psi \cdot \frac{d\theta}{dt}}, \quad \left\{ \cos \psi \cdot \frac{d\theta}{dt} \neq \Omega_1 \right\} \quad \Rightarrow$$

$$\frac{d\theta}{dt} = \frac{\cot \psi \cdot \Omega_1 - \Omega_2}{\cot \psi \cdot \cos \psi + \sin \psi} = \cos \psi \cdot \Omega_1 - \sin \psi \cdot \Omega_2 \quad (19)$$

But the excluding of the term ($d\varphi/t$) by linear combinaton of 2-nd equation, 3-rd equation of system (18), and equation (19) yields

$$\cot \theta = \frac{(\Omega_3 - \frac{d\psi}{dt})}{(\cos \psi \cdot \Omega_2 + \sin \psi \cdot \Omega_1)} \quad \left\{ \tan \psi \neq -\frac{\Omega_2}{\Omega_1} \right\} \quad (20)$$

If we differentiate both the parts of Eqn. (20) with respect to the time-parameter $t$, we could obtain from Eqn. (19) and Eqn. (20)

$$\frac{d(\cot \theta)}{dt} = \left( \frac{\Omega_3 - \frac{d\psi}{dt}}{\cos \psi \cdot \Omega_2 + \sin \psi \cdot \Omega_1} \right)', \quad \Rightarrow$$

$$(-\cos \psi \cdot \Omega_1 + \sin \psi \cdot \Omega_2) \cdot \left( 1 + \left( \frac{\Omega_3 - \frac{d\psi}{dt}}{\cos \psi \cdot \Omega_2 + \sin \psi \cdot \Omega_1} \right)^2 \right) = \frac{d\left( \frac{\Omega_3 - \frac{d\psi}{dt}}{\cos \psi \cdot \Omega_2 + \sin \psi \cdot \Omega_1} \right)}{dt} \quad (21)$$

Equation (21) determines the dynamics of the angle $\psi(t)$, which is depending on the components of angular velocity of asteroid rotation. We should note that the last equation (21) is obviously extremly non-linear ordinary differential equation of the 2-nd order, which could be solved only by numerical methods.

Having solved it in regard to the angle $\psi(t)$, we could obtain the appropriate expression for the angle $\theta(t)$ from Eqn. (20) as well as we could solve properly the 2-nd equation of system (18) in regard to angle $\varphi(t)$.

Let us consider the simplifying case of asymptotical rotations of asteroid $\psi \to 0$ (note that if we additionally assume $\varphi \to 0$ it would mean the case *close to* the planar oscillations and rotations of an asteroid around its center of mass, whose spin axis should be oscillating around velocity vector of asteroid (in its orbital plane); recall that we consider the case $I_1 \geq I_2 \cong I_3$ of the principal moments of inertia. If we additionally consider the case of *negligible harmonic oscillations* (16)-(17) for the absolute magnitudes of the components of angular velocity $\{\Omega_2, \Omega_3\} \to 0$ along with the assumption $\psi \sim 0$, equation (19) could be simplified as below (*as first approximation*):

$$\frac{d\theta}{dt} \cong \Omega_1 \quad \Rightarrow \quad \theta = \Omega_1 \cdot (t - t_0) \qquad (22)$$

so, equation (20) should be updated accordingly, *as first approximation*

$$\frac{d\psi}{dt} + (\cot\theta \cdot \Omega_2)\cos\psi + (\cot\theta \cdot \Omega_1)\sin\psi - \Omega_3 = 0 \quad \Rightarrow$$

$$\frac{d\psi}{dt} + \left(\cot(\Omega_1 \cdot (t-t_0)) \cdot \Omega_1\right)\cdot\sin\psi + \left(\cot(\Omega_1 \cdot (t-t_0)) \cdot \Omega_2\right)\cos\psi - \Omega_3 = 0 \qquad (23)$$

The last Eqn. (23) could be reduced by the change of variables $u(t) = \tan(\psi/2)$ to the appropriate *Riccati* ODE, see example (1.79) in [6], p.305:

$$\frac{du}{dt} = \frac{\left(\cot(\Omega_1 \cdot (t-t_0))\cdot\Omega_2 + \Omega_3\right)}{2}\cdot u^2 - \left(\cot(\Omega_1 \cdot (t-t_0))\cdot\Omega_1\right)\cdot u - \frac{\left(\cot(\Omega_1 \cdot (t-t_0))\cdot\Omega_2 - \Omega_3\right)}{2} \qquad (24)$$

which could be reduced as below in case of $\{\Omega_2, \Omega_3\} \to 0$ (*see Fig.2 just to schematically present the type of such the solution*)

$$\int \frac{du}{u} \cong -\Omega_1 \cdot \int \left( \cot(\Omega_1 \cdot (t-t_0)) \right) dt \quad \Rightarrow \quad \ln u \cong -\ln\left( \sin(\Omega_1 \cdot (t-t_0)) \right) \quad \Rightarrow$$

$$u \cong \frac{u(0)}{\sin(\Omega_1 \cdot (t-t_0))} \quad \Rightarrow \quad \psi \cong 2\arctan\left( \frac{\tan(\psi(0)/2)}{\sin(\Omega_1 \cdot (t-t_0))} \right) \quad (25)$$

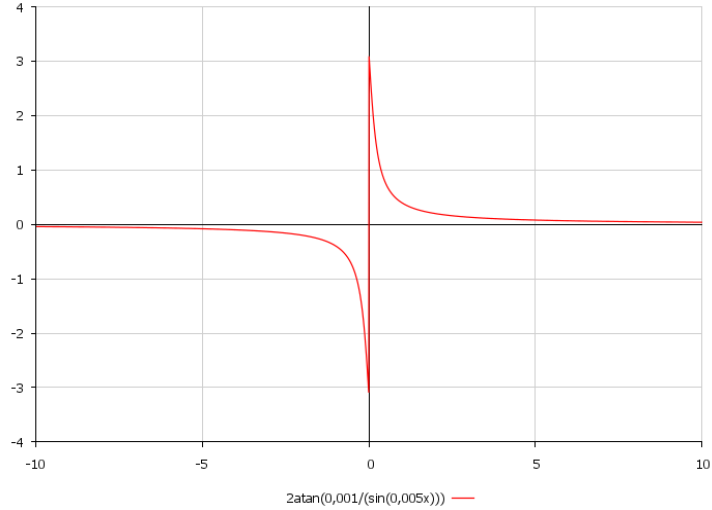

Fig.2. Schematically presented the solution of a type (25).

As we can see from Fig.2, even the asymptotical solutions of Eqs. (23)-(24) reveal their special character of the solutions of *Riccati*-type ODE [6] (for which there exists a possibility for sudden *jumping* of the magnitude of the solution at some meaning of time-parameter *t*). Nevertheless, by the adjusting of the proper initial conditions, such the solution could be obtained to be *negligible* enough indeed, see Fig.3:

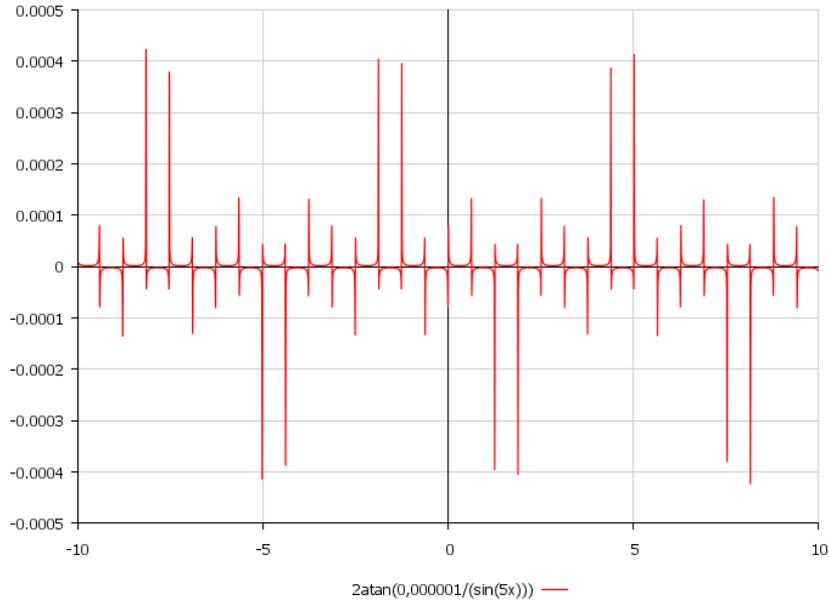

Fig.3. Schematically presented solution (25) of a *negligible* magnitude.

As for the dynamics of angle $\varphi$, we could obviously obtain it from the 1-st of Eqs. (18), Eqn. (19) as well as from the Eqs. (16)-(17) (under assumption $\psi \to 0$):

$$\Omega_1 = \sin\psi \cdot \sin\theta \cdot \frac{d\varphi}{dt} + \cos\psi \cdot (\cos\psi \cdot \Omega_1 - \sin\psi \cdot \Omega_2), \Rightarrow$$

$$\sin^2\psi \cdot \Omega_1 = \sin\psi \cdot \sin\theta \cdot \frac{d\varphi}{dt} - \cos\psi \cdot \sin\psi \cdot \Omega_2, \Rightarrow$$

$$\frac{d\varphi}{dt} \cong \frac{\psi \cdot \Omega_1 + \Omega_2}{\sin(\Omega_1 \cdot (t-t_0))} \cong \frac{2\Omega_1 \cdot \tan(\psi(0)/2)}{\sin^2(\Omega_1 \cdot (t-t_0))}, \quad \left\{\psi \cong \frac{2\tan(\psi(0)/2)}{\sin(\Omega_1 \cdot (t-t_0))}\right\}$$

$$\int d\varphi \cong -2\tan(\psi(0)/2) \cdot \cot(\Omega_1 \cdot (t-t_0)), \tag{26}$$

- it means that rate of changing of angle $\varphi$ is to be periodically changing during the motion of asteroid in its elliptic orbit through Space, see Fig.4.

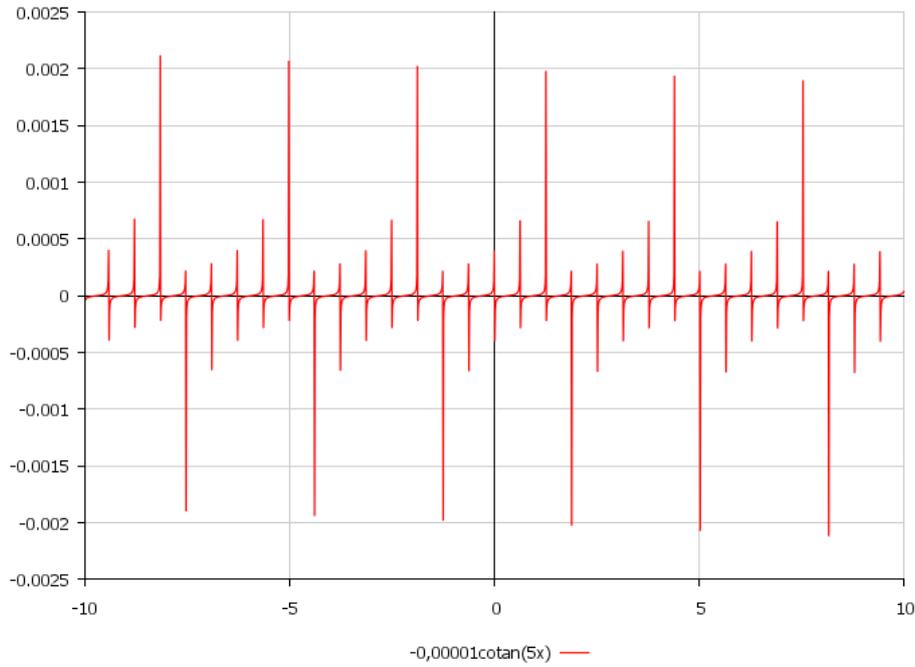

Fig.4. Schematically presented solution (26) of a *negligible* magnitude.

## Appendix A2. The dynamics of the asteroid rotation via *Wisdom* angles.

According to [30], the Euler angles are not suitable for the solving of Eqs. (1) because the resulting equations have a coordinate singularity when the spin axis is normal to the orbit plane (when $\sin\theta = 0$), which could be just the situation under study. A more convenient set of angles has therefore been chosen and is specified relative to an inertial coordinate system $Ox$, $Oy$, $Oz$ which is defined earlier in section Appendix *A*1. Three successive rotations are performed to bring the $Ox'$, $Oy'$, $Oz'$ axes (*which is the frame of reference fixed in the rotating body*) to their actual orientation from an orientation coincident with the $Ox$, $Oy$, $Oz$ set of axes. First, the $Ox'$, $Oy'$, $Oz'$ axes are rotated about the $Oz'$ axis by an angle $\theta_w$. This is followed by a rotation about the $Ox'$ axis by an angle $\varphi_w$. The third rotation is about the $Oy'$ axis by an angle $\psi_w$.

Notation of the angles is chosen the same as that of [20], [29]-[30], but their names have been interchanged with respect to the case of Euler rotations (see Appendix *A*1). In terms of these angles, three angular velocities are [30]:

$$\Omega_1 = -\frac{d\theta_w}{dt}\cos\varphi_w \cdot \sin\psi_w + \frac{d\varphi_w}{dt}\cos\psi_w,$$

$$\Omega_2 = \frac{d\theta_w}{dt}\sin\varphi_w + \frac{d\psi_w}{dt}, \qquad (27)$$

$$\Omega_3 = \frac{d\theta_w}{dt}\cos\varphi_w \cdot \cos\psi_w + \frac{d\varphi_w}{dt}\sin\psi_w.$$

Let us present the *kinematic* equations of Wisdom angles (27) in other form [20]:

$$\frac{d\theta_w}{dt} = \frac{(\Omega_3 \cdot \cos\psi_w - \Omega_1 \cdot \sin\psi_w)}{\cos\varphi_w},$$

$$\frac{d\varphi_w}{dt} = \Omega_1 \cdot \cos\psi_w + \Omega_3 \cdot \sin\psi_w, \qquad (28)$$

$$\frac{d\psi_w}{dt} = \Omega_2 - \frac{d\theta_w}{dt}\sin\varphi_w.$$

The 3-rd equation of system (28) yields:

$$\sin\varphi_w = \frac{\Omega_2 - \dfrac{d\psi_w}{dt}}{\dfrac{d\theta_w}{dt}}, \quad \Rightarrow \quad \cos\varphi_w \frac{d\varphi_w}{dt} = \left(\frac{\Omega_2 - \dfrac{d\psi_w}{dt}}{\dfrac{d\theta_w}{dt}}\right)',$$

which could be transformed, using the 1-st and 2-nd Eqs. (28), accordingly

$$\frac{(\Omega_3 \cdot \cos\psi_w - \Omega_1 \cdot \sin\psi_w)}{\dfrac{d\theta_w}{dt}} \cdot (\Omega_1 \cdot \cos\psi_w + \Omega_3 \cdot \sin\psi_w) = \left(\frac{\Omega_2 - \dfrac{d\psi_w}{dt}}{\dfrac{d\theta_w}{dt}}\right)', \quad \Rightarrow$$

$$\frac{\left(\Omega_2' - \frac{d^2\psi_w}{dt^2}\right)\cdot\frac{d\theta_w}{dt} - \frac{d^2\theta_w}{dt^2}\cdot\left(\Omega_2 - \frac{d\psi_w}{dt}\right)}{\frac{d\theta_w}{dt}} = \Omega_1\cdot\Omega_3\cdot(\cos^2\psi_w - \sin^2\psi_w) + (\Omega_3^2 - \Omega_1^2)\cdot\cos\psi_w\cdot\sin\psi_w$$

$$\Rightarrow \quad \frac{d^2\theta_w}{dt^2} + \left(\frac{\frac{d^2\psi_w}{dt^2} + \Omega_1\cdot\Omega_3\cdot\cos 2\psi_w + \frac{1}{2}(\Omega_3^2 - \Omega_1^2)\cdot\sin 2\psi_w - \Omega_2'}{\Omega_2 - \frac{d\psi_w}{dt}}\right)\cdot\frac{d\theta_w}{dt} = 0 \qquad (29)$$

Equation (29) determines the dynamics of the angle $\theta_w$, which is depending on the angle $\psi_w$ along with the components of angular velocity of asteroid rotation.

As for the dynamics of angle $\varphi_w$, we could obviously obtain from the 2-nd of Eqs. (28) as well as from the 1-st equation of Eqs. (16):

$$\frac{d\varphi_w}{dt} = \Omega_1\cdot\cos\psi_w + \Omega_3\cdot\sin\psi_w\,.$$

## Appendix A3. How the *Wisdom* angles depend on the *Euler* angles.

The last but not least, we should demonstrate how the the *Euler* angles could be transformed to the *Wisdom* angles. According to [29], such an extremly non-linear connexion is presented below

$$\tan\psi_w = \tan\varphi\cdot\sin\psi,$$

$$\tan\varphi_w = \frac{\sin\varphi\cdot\cos\psi}{(\cos\varphi/\cos\psi_w)}, \qquad (30)$$

$$\tan\theta_w = \frac{\cos\theta\cdot\sin\psi + \sin\theta\cdot\cos\varphi\cdot\cos\psi}{-\sin\theta\cdot\sin\psi + \cos\theta\cdot\cos\varphi\cdot\cos\psi}.$$

As for the components of the approximate solution (22), (25)-(26), recall that we used the simplifying assumption for asteroid rotation $\psi \to 0$ at resolving of Eqs. (18)-(20) (along with the assumption for angle $\varphi \to 0$ as well). As we can see from the equations (30), such an assumptions yields

$$\psi_w \cong 0,$$

$$\tan \varphi_w \cong \tan \varphi \to 0, \qquad (31)$$

$$\cot \theta_w \cong \frac{-\tan \theta \cdot \sin \psi + 1}{\sin \psi + \tan \theta} \cong \cot \theta - \sin \psi.$$

Thus, the only essential difference between two presentations of the approximate solution (22), (25)-(26) (in *Euler* angles versus *Wisdom* angles) is that the component $\theta_w$ should oscillate (see Eqs. (31)) near the meaning:

$$\theta = \Omega_1 \cdot (t - t_0)$$

- where the maximal magnitude of oscillations is given by the expression (25).